RADIATION AS A CONSTRAINT FOR LIFE IN THE UNIVERSE

Ximena C. Abrevaya[1], Brian C. Thomas[2]

1. Instituto de Astronomía y Física del Espacio (UBA–CONICET), Buenos Aires, Argentina
2. Washburn University, Topeka, KS, United States

Abstract

In this chapter, we present an overview of sources of biologically relevant astrophysical radiation and effects of that radiation on organisms and their habitats. We consider both electromagnetic and particle radiation, with an emphasis on ionizing radiation and ultraviolet light, all of which can impact organisms directly as well as indirectly through modifications of their habitats. We review what is known about specific sources, such as supernovae, gamma-ray bursts, and stellar activity, including the radiation produced and likely rates of significant events. We discuss both negative and potential positive impacts on individual organisms and their environments and how radiation in a broad context affects habitability.

1. INTRODUCTION

There are several factors that can constrain the existence of life on planetary bodies. To determine the possibility of existence and emergence of life, it is essential to consider astrophysical radiation which itself can be a constraint for the origin of life and its development. Additionally, the radiation received by the planetary body and the plasma environment provided by the parent star play a crucial role on the evolution of the planet and its atmosphere. Therefore, radiation can determine the conditions for the origin, evolution, and existence of life on planetary bodies.

2. TYPES OF RADIATION

Several types of radiation are relevant to life in the universe. The word "radiation" itself may first need some definition. We use this term very broadly, to cover both electromagnetic radiation and energetic particles. The electromagnetic radiation of interest to us is the higher energy end of the spectrum — gamma-ray, X-ray, and ultraviolet. Gamma-ray and X-ray forms of light are ionizing, and along with UV, have the potential to destroy or damage essential biological molecules of life "as we know it," including DNA and proteins. Energetic particles that may cause damage include electrons, protons, neutrons, and muons. The energy of these particles is determined mainly by their kinetic energy. A muon is an elementary particle similar to an electron, but with a greater mass. Muons are highly penetrating and ionizing, but not very much is known about their biological effects. Unlike gamma-rays or neutrons, or even electrons, muons are not produced in most artificial sources of radiation, and so their biological effects have not been studied (Atri and Melott, 2011; Rodriguez et al. 2013). They are normally assumed to have effects similar to electrons, but could be more severe due to greater penetration (Fig. 1).

High-energy electromagnetic radiation is produced by many different processes. UV light is produced by blackbody emitters with sufficiently high temperatures, including our own Sun. In general, the ultraviolet region of the electromagnetic spectrum can be subdivided into bands. These subdivisions are arbitrary and can differ depending on the discipline involved (Diffey, 1991), but in biology it is possible to distinguish three main bands: UVA (400–315 nm), UVB (315–280 nm), and

UVC (280–100 nm) and additional bands for shorter wavelengths as VUV (200–10 nm) or EUV (121–10 nm). Gamma- and X-rays can be produced by radioactive decay of certain elements, by electron-positron pair annihilation, inverse-Compton scattering, some intra-atom electron transitions, and Bremsstrahlung radiation.

Charged particles (e.g., electrons, protons) can be accelerated to high energies by various processes, especially involving plasma shocks and interactions with magnetic fields. (For readers interested in more of the physics involved, we suggest Rieger et al. (2007) and Balogh and Treumann (2013).) Neutrons and muons are generated by nuclear reactions. Neutrons may be ejected from nuclei that radioactively decay (e.g., a Uranium nucleus). Both neutrons and muons can be generated by nuclear reactions of atomic nuclei with high-energy "primary" protons that enter a medium such as a planetary atmosphere. The primary protons induce a so-called air shower of secondary particles, which includes electromagnetic and particle constituents, including neutrons and muons. In addition, helium nuclei (termed "alpha particles") and electrons are produced in some radioactive decay processes. Electrons with relatively high energy are also found in the magnetospheres of planets with significant magnetic fields. This is the case for some terrestrial planets (e.g., Earth) as well as giant planets. The moons of giant planets may experience significant irradiation due to the planet's magnetospheric electrons; this is the case for Jupiter's four largest moons, for instance. In this case, how- ever, the electron radiation is not very penetrating, so some shielding by ice/rock will prevent significant impacts below the surface.

## 3. SOURCES OF HIGH-ENERGY RADIATION

### 3.1 STELLAR EMISSIONS

Stars are of course sources of visible light, but they also emit UV, X-ray, and even gamma-ray light. The emission of stellar radiation depends on their surface temperature and also on their activity. Therefore, emission is related to the evolution and spectral type of the star and it can be highly variable. A star's surface temperature determines its blackbody emission. A high-mass star on the main sequence will have a high surface temperature and emit a significant amount of UV. However, from the perspective of habitability, such high-mass stars are short-lived (as short as a few million years), probably too short- lived to host habitable planets (Turnbull and Tarter, 2003). Stars that are of most interest for habitable planets include those similar to the Sun's mass (types G and K), but also those of much lower mass (type M) (Tarter et al., 2007; Kopparapu et al., 2013).

The Sun emits enough UV light to be problematic for planets without a UV shield, such as ozone in the atmosphere. Lower mass stars, on the other hand, do not emit very much UV, but are more active, with frequent energetic flares. These flares themselves are sudden and energetic explosive events and emit UV and X-ray (and possibly some gamma-ray) light. They originate in magnetic processes affecting all the layers of the stellar atmosphere (photosphere, chromosphere, and corona), which heat the stellar plasma and accelerate its protons, electrons, and heavy ions, to velocities near the speed of light. Flare emissions (usually being several magnitudes higher compared to the quiescent state) undergo interactions with planets and it is not well-understood whether it could be lethal or unfavorable for life. In general, it is known that the strongest stellar flares exceed the strongest solar ones by a factor of 100 in X-ray and EUV flux. The quiescent X-ray and EUV radiation of young stars are up to a factor of 1000 higher than on the present-day Sun (Guinan and Ribas, 2002; Ribas et al., 2005). They also are likely to eject plasma through processes similar to those that produce Coronal Mass Ejections on our own Sun. CMEs and flares represent important sources of both electromagnetic and particle radiation. The intermittent nature of the emission from low-mass stars may present more of a hazard than a steady background emission (such as UV from a Sun-like star), since it may be harder for life to adapt to the varying levels of radiation, as opposed to a more constant value (e.g., Ayres, 1997; Gershberg, 2005; Scalo et al., 2007).

Over a star's lifetime, its radiation emission changes. A young star tends to be less luminous overall, but often more active, producing more frequent and more intense flare/CME events. As a star ages its luminosity slowly increases, but the activity tends to decrease; this decrease is more pronounced for stars of higher mass, while low-mass (e.g., M-dwarf) stars continue to be highly active.

3.2 STELLAR EXPLOSIONS

Explosions on the scale of whole stars fall into a couple of major categories: individual stars that explode and pairs of stars that interact leading to explosions. These events are usually categorized by how they are observed. A supernova is typically observed as a rapid brightening in visible light. Observations can also be made in UV and, for a few cases, there are observations in X-ray and gamma-ray; the data is limited in these wavebands due to the lower luminosity, but also the relative lack of observational equipment.

Supernovae are categorized by features in their light curve (the variation in luminosity with time) and the strength of the hydrogen absorption lines in their spectra. Type I events have a sharp increase in luminosity followed by a steady, gradual dimming, and show little to no H absorption, while Type II have a sharp increase in luminosity followed in most cases by a plateau lasting a few months and then a gradual dimming, and show stronger H lines; each type also has subtypes determined by other details in the spectrum. For a recent review see Hillebrandt (2011).

Broadly, Type II events are explosions of individual high-mass stars that undergo core collapse. This progenitor is also responsible for Type IB and Type IC supernovae, but in these cases H absorption is weak. Type IIL, IIP, IIN, and IIB are defined by differences in the spectra, except that a Type IIL does not show the light curve plateau that Type IIP supernovae do.

A Type Ia supernova, on the other hand, is thought to be the explosion of a white dwarf that has accreted matter from a companion (larger, main sequence, or giant) star. In this model, the white dwarf is near the critical mass of 1.4 solar masses (the Chandrasekhar limit), which is the most mass that can be supported by the electron degenerate matter that makes up a white dwarf. When more mass is accreted, the star collapses and explodes.

All supernovae produce visible and UV light and likely all produce higher energy light as well, though observations are limited. Gamma-rays emitted from supernovae are the result of radioactive decay of certain elements that are synthesized in the explosion process (see, for instance, Karam, 2002a,b). Supernovae emit much of their energy in neutrinos, almost massless elementary particles, which interact so weakly as to pose no threat to organisms (Karam, 2002b).

Supernovae also produce an ejecta blast wave that propagates outward. These blast waves form "remnants" that are visible for some time after the explosion and inject the progenitor and synthesized material into the interstellar medium. In addition, the shock front accelerates protons to high energies, producing at least a portion of the cosmic rays observed on Earth, which would also be present for most other habitable planets. An exception may be moons of giant planets which could be shielded by their host planet's strong magnetic field (in which case, however, those moons would be subject to the magnetospheric radiation of the planet, as noted above).

Another category of stellar explosions, again defined by how they are observed, is gamma-ray bursts (GRBs). As the name implies, they are observed initially as a "burst" of gamma radiation, which is followed by emission in lower-energy wavebands, all the way through radio. For an excellent review of GRBs, see Gehrels et al. (2009). These bursts fall into two subcategories, "long" and "short," defined by the duration of the gamma-ray emission. Long GRBs are of order 10's of seconds, while short GRBs are about 1 s or less (in both cases referring only to the gamma emission; the "afterglow" in other wave- bands may last much longer). The two types also show a difference in their spectra, with long GRBs having "softer" spectra, dominated by lower-energy gamma-rays (with a spectral peak around 100–200 keV), and short GRBs having

"harder" spectra, with greater emission of high-energy gamma-rays (with a spectral peak closer to 1 MeV).

The progenitors of long GRBs are most likely individual stars that explode as core-collapse super- novae and are situated such that they launch an intense "jet" of material along their rotation axis which happens to be pointed at Earth, leading to the burst of high-energy light observed. The fact that the emission is strongly "beamed" allows for what may be a fairly normal supernova explosion to be observed as such an intense blast. While this scenario is the most widely accepted model, the full picture may be more complicated. (For a good review of GRBs, see Kouveliotou et al. (2012).) Short GRBs, while also thought to emit radiation along a jet, are most likely the result of the merger of two compact objects, such as neutron stars or black holes.

Other short-term stellar events also produce high-energy radiation, but are of low enough intensity as to not be significant on large scales. These include "soft gamma repeaters" thought to be powered by "magnetar" stars that periodically emit lower energy gamma-rays, but with relatively low luminosity.

Black holes are also a source of high-energy radiation, particularly X-rays and energetic protons, but only if they are actively accreting matter. This is most likely in the case of supermassive black holes associated with active galactic nuclei (AGNs). Emission from stellar mass black holes is rare enough and of small enough luminosity to not be significant from the point of view of habitability. AGN, on the other hand, may be significant, when the black hole is particularly active, and could have an effect on much of their host galaxy, primarily through accelerating particles to high energies, thereby increasing the background cosmic ray flux.

4. EFFECTS

In the previous sections, we described the main astrophysical sources of radiation in the universe and the different types of radiation that can be derived from them. Two main factors determine the effect of radiation on habitability: the total energy received by a given habitat and the "hardness" of the radiation (where hardness refers to the relative amount of higher- to lower-energy photons or particles received from the source). That means that the effects of radiation on life will depend in fact on the kind of radiation (electromagnetic or particle and their energy), the amounts of radiation (dose or fluence), and the capability of the living beings to cope with radiation.

Biologically, damaging radiation could reach the surface of the planet, depending on the existence of a magnetic field and the presence of an atmosphere. Magnetic fields can shield the surface from charged particles, depending on the strength of the field and "rigidity" (a combination of momentum and charge) of the particles. An atmosphere can protect from both particle and electromagnetic radiation depending on the energy of the radiation and thickness of the atmosphere (Dartnell, 2011).

We can consider effects on life as being either direct or indirect. Direct effects involve the interaction of radiation directly from the event with biological material (cells, prebiotic molecules); mean- while indirect effects are those related to the interaction of the radiation with the environment (atmosphere), therefore favoring or limiting the possibility of life to arise and evolve (Abrevaya, 2013).

A fair amount of work has been done on the subject of astrophysical ionizing radiation and life. We cite much of that work below and also refer interested readers to the excellent reviews by Horneck et al. (2010), Olsson-Francis and Cockell (2010), and Dartnell (2011).

4.1 DIRECT EFFECTS

In general, radiation can be very harmful and even lethal to living beings, as it is capable of damaging DNA and other cellular components through different kinds of mechanisms. If we consider a planet with an atmosphere and magnetic field, UV radiation will be capable of reaching the surface, as well as muons and neutrons if sufficient energetic particles are incident at the top of the atmosphere.

In the case of UV, the most damaging effects are exhibited through direct interaction of UV photons with essential macromolecules such as DNA or proteins. As these molecules have a maximum of absorption of UV radiation at 260 nm and 280 nm, respectively, these effects are seen at UVC (100–280 nm) and UVB wavelengths (280–315 nm). The predominant kinds of damage on DNA are chemical modifications such as cyclobutane pyrimidine dimers (CPDs) and (6–4) photoproducts (6–4PPs). DNA single-strand and double-strand breaks can also be induced by UV, but these are produced as a consequence of failures during the DNA repair steps of CPDs and 6–4PPs, as was described in Bonura and Smith (1975a,b) and later by Bradley (1981).

Other kinds of damage are produced by indirect mechanisms, for example, at longer wavelengths as UVA (315–400 nm) where the absorption of DNA and proteins is null or very weak. In this case, free radicals such as reactive oxygen species are generated during the radiolysis of water molecules. The hydroxyl radical (OH) is the main damaging species producing a plethora of DNA lesions in the form of chemical modifications (e.g., 8-hydroxyguanine, DNA-protein cross-links) (for more details see Kielbassa et al., 1997 and references therein).

Other cellular components can also be damaged by UV, such as proteins. Oxidation of prokaryotic proteins during irradiation was documented for different microorganisms (Daly et al., 2007; Qiu et al., 2006). It was also suggested that UV radiation can damage membrane proteins with the concomitant leakiness of membranes (Koch et al., 1976). Membrane damage was also documented for microorganisms exposed to the 200–400 nm UV range (Fendrihan et al., 2009).

UV is also capable of inhibiting metabolism, enzymatic activity, and several cellular processes in general, such as photosynthesis (Sinha et al., 1995; Renger et al., 1989; Neale et al., 1998; Neale and Thomas, 2016).

From the experimental point of view, few works analyzed the effects of stellar UV radiation on life considering planets orbiting habitable stars (G, F, K, and M-type stars). Fendrihan et al. (2009) exposed halophilic archaea to several UV doses over a wavelength range of 200–400 nm to simulate the Martian UV flux. Cells that were embedded in halite showed survival under UV exposure doses as high as $10^4$ kJ m$^{-2}$ (exposure at Earth's surface today is around 3-4 kJ m$^{-2}$). Cockell et al. (2005) also exposed dried monolayers of *Chroococcidiopsis sp.* 029, a desiccation-tolerant, endolithic cyanobacterium, to a simulated martian-surface UV and visible light flux, also equivalent to the worst-case scenario for irradiation conditions on the Archean Earth. They have found loss of viability after 30 min of exposure.

The probability of survival of radiation-tolerant microorganisms (halophilic archaea) was evaluated considering flare activity from the dM star EV-Lacertae (EV Lac, Gliese 873, HIP 112460) taking the UVC region (254 nm). Microorganisms survived the exposure to irradiation conditions (Abrevaya et al., 2011a). The same UV-resistant profiles were observed in experiments simulating radiation of the interplanetary environment or exposed in the low Earth orbit, where microorganisms have been exposed to EUV (e.g., Mancinelli et al., 1998; Abrevaya et al., 2011b; Mancinelli, 2015). Other works have analyzed potential effects on life of stellar UV radiation, but they are only based on theoretical modeling and do not consider their effects on microorganisms but on isolated DNA molecules (Cockell, 1998, 1999; Cockell et al., 2005; Scalo and Wheeler, 2002; Rontó et al., 2003; Segura et al., 2003, 2010; Cockell and Raven, 2004; Buccino et al., 2007; Cuntz et al., 2010; Rugheimer et al., 2015).

At wavelengths shorter than UV, the effects of X-rays and gamma-rays are also well known. In general, direct action on the DNA molecule produces both DNA single-strand and double-strand breaks. Additionally, damage through indirect mechanisms as free radicals by radiolysis of water molecules is generated. There is no direct experimental data on the effects of this kind of radiation in the planetary context. Theoretical modeling has made predictions concerning the effects of radiation on the Earth's biosphere and revealed the biological importance of UV-flashes from GRBs delivered to the surface of the Earth, considering different present and prehistoric atmospheres (Galante and Horvath, 2007; Martín et al., 2009, 2010; Horvath and Galante, 2012).

On the other hand, since the "flash" from a GRB lasts at most 10s of seconds, this may have only a small impact on the biosphere. It is likely that the more important aspect, in the long run, is severe depletion of stratospheric $O_3$, caused by the formation of odd-nitrogen oxides after ionization induced by high-energy photons and cosmic rays (in the case of nearby supernovae). Thomas et al. (2005), for instance, estimated an increment in the DNA damage of up to 16 times the normal annual global average, which may be lethal for microorganisms such as phytoplankton. On the other hand, the bio- logical impacts of increased UV following a GRB or similar event are complicated and depend on the particular organism or impact considered (Thomas et al., 2015). For two important (modern day) oce- anic primary producers, Neale and Thomas (2016) found only a small impact on productivity. However, this study was limited in that it modeled only relatively short-term impacts and much remains to be learned about the long-term effects, including the level of mortality under post-GRB-type conditions.

Based on anticipated effects of reduced $O_3$, it has been argued that GRBs are likely to have impacted the Earth during the last billion years and could be responsible for mass extinctions (Melott et al., 2004; Melott and Thomas, 2009, 2011).

If we consider the space radiation environment, high-energy charged particles are present and they can interact at multiple scales with biological structures. Additionally, they can produce secondary particles capable of interacting with biological material. This kind of radiation should be distinguished from X-rays or gamma-rays as their deposition energy is done through a different mechanism along a "linear" track. Therefore, this produces distinguishable biological effects, different from those generated by other kinds of radiation, as particles can induce instantaneous damage, which is not compatible with repair mechanisms on cells, for example, when damaging molecules such as DNA. A detailed description of this phenomenon can be found in Nelson (2003). Some biological effects of low-energy particle radiation are also described in Yang et al. (1991).

Taking into account charged particles in an astrobiological context, Paulino-Lima et al. (2011) replicated charged particles under laboratory conditions to simulate solar wind. The radio-resistant micro- organism *Deinococcus radiodurans* was exposed to electrons, protons, and ions to test its probability of survival. The results indicated that low-energy particle radiation (2–4 keV) had no significant effects on the survival of this microorganism, even if the microorganisms were irradiated with an equivalent fluence of 1000 years of exposure at 1 AU. However, as the authors mention, the effect of high-energy ions as those we could find in solar flares (200 keV) could have more deleterious effects on microbial cells, with estimated 90% cell inactivation, considering a distance of 1 AU and several flare events in one year.

It should be noted, however, that life on Earth evolved to cope with radiation as cells have developed different strategies that allow repair or prevent damage. Different DNA repair systems depending on specific enzymes exist in all life forms "as we know it" and are necessary to recognize and rebuild the injured sites, to prevent cell death. These processes are diverse from the point of view of mechanisms, but globally are highly conserved from prokaryotes to eukaryotes (and also including some viruses such as bacteriophages) (Cromie et al., 2001). One of the most unique and relevant features in the radiation-resistant microorganism par excellence, *D. radiodurans*, is its extremely powerful DNA repair mechanism (e.g., Cox and Battista, 2005). Several hypotheses have been suggested to explain the evolution of DNA repair and can be found in O'Brien (2006).

During biological evolution, living beings also developed other physiological strategies not only to repair DNA damage, but also to prevent it. Pigments, for example, such as melanin (Brenner and Hearing, 2008; Cordero and Casadevall, 2017), can act as a radiation shield, in particular for UV. Scytonemin, a sheath pigment in cyanobacteria, was found to protect these microorganisms against UVC radiation (Garcia Pichel et al., 1992; Dillon and Castenholz, 1999). Carotenoid pigments have also shown to protect microorganisms from UV. In fact, a positive correlation between the presence of carotenoids and resistance to radiation in bacteria was already documented several decades ago (Mathews and Krinsky, 1965). Moreover, carotenoids could have a role in DNA repair mechanisms such as photoreactivation or act as protective agents against the effects of free radicals such as hydrogen peroxide (Shahmohammadi et al., 1998). A detailed review of UV screening com- pounds and its relevance can be found in Cockell and Knowland (1999).

In haloarchaea, high intracellular concentrations of KCl seem to also provide protection against radiation through interaction with free radicals (Kish et al., 2009). Other radio-resistant microorganisms such as *D. radiodurans* showed that high intracellular Mn/Fe ratio combined with desiccation contributes to ionizing radiation resistance (Paulino-Lima et al., 2016). Also, physiological mechanisms such as polyploidy present in haloarchaea seem to provide advantages against radiation damage (Breuert et al., 2006).

Additionally, highly resistant structures such as bacterial spores (dormant structures produced by some bacteria that are formed in response to adverse environmental conditions) have also been shown to offer effective protection against the effects of UV radiation. Results obtained by Risenman and Nicholson (2000) indicate that the spore coat in *Bacillus subtilis* endospores is necessary for spore resistance to environmentally relevant UV wavelengths. Spores have also been shown to be 10- to 50-times more resistant to UV than growing cells and also more resistant to gamma radiation than cells during the growing state (Nicholson et al., 2000, 2005). Different kinds of photoproducts can be gen- erated in spores by UV irradiation than those acquired when *B. subtillis* is in its growing state (Setlow, 2006). A summary can be found in Horneck et al. (2014).

In addition to physiological mechanisms that provide protection against radiation, the habitat where life forms exist and develop can be also particularly protective, for instance, in the cases of endolithic microorganisms living inside rocks (a detailed description of different kind of endoliths can be found in Golubic et al., 1981) and evaporitic environments. In addition to the obvious case of shielding from UV by opaque rock materials, haloarchaea inhabiting fluid inclusions of halite crystals have also been shown to be protected, as these crystals absorb short UV wavelengths and reemit them at longer, less damaging wavelengths (Fendrihan et al., 2009). In a series of works, Horneck et al. (2001) and Rettberg et al. (2002, 2004) also showed that thin layers of clay, rock, or meteorite material are successful in UV-shielding.

Aquatic ecosystems can also provide shielding from UV depending on the optical properties of the water that control light penetration, which is influenced by dissolved and suspended organic material (Diffey, 1991). Vertical mixing has also been found to be an important factor (Huot et al., 2000).

This chapter is focused mainly on negative effects of radiation, but astrophysical radiation has likely had positive effects for life, especially in the context of prebiotic molecules. UV radiation, for instance, could have played an important role during the polymerization of the first prebiotic organic molecules (Dauvillier, 1947; Ponamperuma et al., 1963; Sagan and Khare, 1971; Sagan,

Abrevaya & Thomas - Radiation as a Constraint for Life in the Universe                                         Page 9 of 29

1973; Pestunova et al., 2005). Ranjan et al. (2017) determined the UV environment on prebiotic Earth-analog planets orbiting M dwarfs such as the recently discovered Proxima Centauri, TRAPPIST-1, and LHS 1140. They obtained dose rates to quantify the impact of different host stars on prebiotically important photoprocesses. According to the results obtained in this study, M-dwarf planets have access to 100-1000 times less bioactive UV fluence than the young Earth. Therefore, it is unclear whether Earth-like planets orbiting M-dwarfs could host UV-sensitive prebiotic chemistry that may have been important to abiogenesis on Earth (e.g.: pyrimidine ribonucleotide synthesis). However, it is unknown if transient elevated UV irradiation due to flares may suffice. Experimental work under laboratory conditions is needed in order to constrain all these possibilities. Atri (2016b) has proposed that cosmic rays could provide energy for existing subsurface radiolysis-powered life. In general, ionizing radiation could have had an important role in the origin of life and is relevant for the generation of habitable planetary environments (Dartnell, 2011).

In the context of biological evolution, other positive effects can be considered if the radiation doses are nonlethal for microorganisms. In this case, they could induce mutations increasing the genetic variability, thus providing new raw material for all sorts of selective pressure. For example, UV can act as a selective pressure itself, leading to the appearance of organisms adapted to live under UV stress, such as those with pigments (Scalo and Wheeler, 2002; Wynn-Williams et al., 2002). It is also postulated that UV radiation could have influenced protistan evolution (Rothschild, 1999).

4.2 INDIRECT EFFECTS

Indirect effects can be seen through the interaction of radiation with the atmosphere. Life on Earth is currently shielded from most ionizing radiation from space. The atmosphere of the present Earth (which started to increase its levels of oxygen around 2.5 Gyr ago) is thick enough to screen out high-energy photons (gamma- and X-rays), $O_2$ absorbs short-wavelength UV (UVC), and ozone in the middle atmosphere absorbs most UV between 200 and 350 nm (the biologically damaging UVB). In the opposite way, primitive Earth, which had a different atmospheric composition (anoxygenic atmosphere), was unable to shield the surface of the planet from the effects of UV radiation through $O_3$, but may have had instead "hazy" conditions that could have reduced the UV transmission (Wolf and Toon, 2010).

Stellar X-rays could affect the atmospheric evolution and the chances for life to emerge (Kulikov et al., 2007; Lammer et al., 2008). Theoretical modeling has shown that this radiation is capable of dissociating $N_2$ and $O_2$ in the atmosphere, releasing important quantities of very reactive species (atomic nitrogen and oxygen) which leads to the formation of nitrogen oxides that act as catalyzers of ozone dissociation, and therefore, increase the irradiation of the planet's surface with stellar UV radiation, among other important effects (Martı́n et al., 2010).

Similarly, high-energy charged particles (cosmic rays, mainly protons) interact with air molecules high in the atmosphere. On the other hand, those interactions lead to "showers" of secondary particles, some of which can be penetrating and damaging, in particular neutrons and muons, depending on the altitude considered. Charged particles with energy below about 10 GeV are deflected by Earth's large-scale magnetic field. Particles with energy below about 1 GeV are mostly deflected by the Sun's field, but that shielding varies with solar activity (more shielding when the Sun is more active). For a review of the effects on terrestrial life by cosmic rays, see Atri and Melott (2014).

On Earth then, life is mainly protected from direct effects of ionizing radiation by a thick atmosphere and large-scale magnetic field. In contrast, Mars has a thin atmosphere and no large-scale magnetic field. Smith et al. (2004a,b) and Smith and Scalo (2007) performed detailed computations of radiative transfer of high-energy photons and found that the surface of Mars would be exposed to a substantial fraction of any incident gamma radiation, while X-rays are effectively blocked. Due to the lack of a large-scale magnetic field, a planet like Mars is exposed to charged particle radiation of all energies. While the atmosphere will shield the surface to some extent, there will still be a significant flux of damaging primary and secondary charged particle radiation at the surface (Dartnell et al., 2007; Pavlov et al., 2012).

Moons around giant planets are generally too small to hold a significant atmosphere or have a large-scale magnetic field. The surfaces of Europa and Enceladus, for instance, are exposed to any incident photons. A moon's host planet can have a strong magnetic field, which, if the moon is sufficiently within that field, provides protection from high-energy cosmic rays, but will at the same time subject the moon to magnetospheric ions and electrons with energy up to tens of MeV (Cooper et al., 2001). However, a few hundred meters of ice and rock are effective shields against both high-energy photons and charged particles, so habitats existing sufficiently deep under the surface of such moons should not be affected even by the most intense irradiation from outside.

While thick, Earth-like atmospheres protect life on the surface from direct radiation effects, that life may still experience increased ionizing radiation during rare but intense high-energy astrophysical events. As noted above, high-energy protons (with energy above a few GeV) generate "showers" of secondary particles. For life around sea-level, energetic muons are the greatest threat. These "heavy electrons" can penetrate several hundred meters of rock, ice, or water and damage biological material. An enhancement of cosmic rays due to, for instance, a nearby supernova can increase the background muon radiation level by several times (depending on factors such as the distance to the supernova; Thomas et al. 2016), lasting hundreds to thousands of years, due to the slow diffusion of charged particles through interstellar space.

In addition, a thick atmosphere can "redistribute" the energy of high-energy photons (gamma- and X-rays) to UV photons, increasing the UV radiation at the surface for the duration of a gamma-ray event (Smith et al., 2004a,b; Smith and Scalo, 2007).

Finally, thick atmospheres can experience an increase in ionization due to both high-energy photons (gamma- and X-rays) and high-energy charged particles (above a few MeV), with higher-energy radiation affecting the atmosphere at lower altitudes. This ionization in an $N_2$-$O_2$ dominated atmosphere can lead to production of nitrogen oxides that catalytically destroy ozone, leading to increased penetration of UV from the host star (Thomas et al., 2005, 2015). This indirect irradiation, in fact, appears to be the most significant effect for Earth-like planets following short duration, high-energy ionizing photon events such as GRBs.

We now summarize what is known about the impacts of specific sources. In all cases, the severity of impacts depends on two main factors: (1) the total energy received, with more energy meaning greater impact and (2) the "hardness" of the radiation spectrum, with a "harder" spectrum having relatively greater flux of high-energy particles/photons, which tend to have a greater impact than lower energy particles/photons. Different types of event (SNe, GRBs, stellar activity) will have different spectra and total luminosity. The received energy depends on the intrinsic luminosity and the distance from the event (except in the case of a planet exposed to its host star's radiation, in which case the distance is negligible). The intensity of radiation decreases with the square of the distance in general, but the dependence may be more complicated for charged particles, which have significant interactions with magnetic fields in the Galaxy that cause diffusive instead of ballistic motion from the source.

GRBs are the simplest source to consider. All GRBs are relatively short in duration, ranging from tens of seconds to fractions of a second. They deliver a burst of high-energy photons, but do not appear to generate charged particle (cosmic ray) flux (Aartsen et al., 2016), at least at the highest energies ($10^{18}$ eV or more). On the other hand, long duration GRBs are known to be associated with supernovae, which are sources of cosmic rays. For planets with thick atmospheres, the high-energy photons lead to redistributed UV radiation at the surface, but this persists only as long as the gamma- and X-rays are incident on the atmosphere, so the effect is quite short-lived (Martín et al., 2009; Peñate et al., 2010). Longer-term atmospheric chemistry effects occur following the ionization induced by the gamma-/X-rays. For planets with significant $O_2$, the chemistry changes lead to destruction of the ozone shield that is naturally present in the middle atmosphere of planets with $O_2$ and a stellar UV flux (Thomas et al., 2005). The destruction of $O_3$ then leads to unusual increases in stellar UV irradiance at the planet's surface and into the first 100 m or so of bodies of water, depending on their clarity (Peñate et al., 2010; Thomas et al., 2015). Overall, the depletion of $O_3$ can last for years to a decade. While there are two categories of GRB, both have essentially the same effect. Short GRBs have a harder spectrum but generally lower luminosity, while long GRBs have a softer spectrum but higher luminosity. Overall, they have very similar effects.

Supernovae are a more complicated source. First, they emit high-energy photons, which travel directly from the source with a $1/r^2$ intensity dependence. The photons are for the most part in the X-ray range and lower, with emission lasting on the order of months. The X-rays will have effects similar to the photon radiation from a GRB, with again the most important result being the depletion of $O_3$. The photons are not high enough in energy (above about 100 keV) to lead to redistributed UV as in the case of a GRB.

Supernovae also accelerate protons in the explosion blast wave. These protons travel outward from the SN ahead of, with, and behind the ejected stellar material. Charged particles follow more complex paths in regions of space with magnetic fields present. Lower-energy particles are more strongly affected and may take many thousands of years longer than the photons to arrive. Higher-energy protons will take a more direct path. If the space in between the SN and the receiving planet is essentially empty of material and magnetic field, then the travel will be more direct and the protons may arrive within a few hundred years of the photons (Kachelrieß et al., 2015).

The accelerated protons will have two main impacts on a planet. First, they will cause ionization in a thick atmosphere, in essentially the same way as high-energy photons. This can lead to depletion of $O_3$, but depends strongly on the spectrum of the received protons. Harder spectra (with more of the higher energy particles) generate ionization closer to the ground and may therefore "miss" the ozone, which is concentrated in the middle atmosphere. However, high-energy protons generate showers of secondary particles, as discussed above, and these secondaries (especially muons) can be damaging at the surface and under hundreds of meters of water, ice, and rock. This is likely to be the most significant biological impact, since ozone depletion is likely to be associated mostly with the photons, which have a duration of months, while the high-energy proton flux will lead to increased biological damage for thousands of years.

For the case of a SN, the presence of a planetary magnetic field is generally not relevant, since the accelerated protons are of high enough energy to be only minimally affected (if at all) by the planet's magnetic field, unless it is much stronger than the present-day Earth's. This is true for isolated planets with their own magnetic fields as well as for moons of giant planets, which will be shielded by their host planet's field from most cosmic ray protons, but may not be shielded from the harder spectrum of pro- tons received from a nearby supernova.

Stellar activity is most significant for close-in planets around lower mass (M type) stars (for an excellent collection of work on this topic see Lammer and Khodachenko, 2015). These stars are more active and the habitable zone is relatively close to the star (due to their low luminosity), meaning that a potentially habitable planet is more directly and more frequently exposed to radiation from stellar flares and CMEs. The relevant radiation in this case is mainly UV and protons. This radiation will mainly affect the atmosphere (see, e.g., Segura et al., 2010; Tabataba-Vakili et al., 2016). The protons will be of too low energy to generate significant showers of secondary particles, therefore not increasing the surface radiation significantly (Atri, 2016a). Another threat to habitability in this environment is atmospheric mass loss due to UV flux and the plasma stellar wind (see for instance Zendejas et al., 2010; See et al., 2014).

There may be some danger for planets located in a galaxy with an active supermassive black hole at its center (an AGN). AGN produce high-energy light (i.e., X-rays) and accelerate protons to very high energies. This may increase the background cosmic ray flux in a fairly steady way for as long as the black hole is active. This could put a constraint on habitability, but on the other hand, a steady enhancement could lead to greater radiation resistance adaptation.

5. RATES

The frequency of "dangerous" ionizing radiation events is relevant to their impact on habitability. Estimating such rates depends on a number of factors. First, as noted above, the most important parameters for determining impact are the total energy received and the hardness of the radiation spectrum. Details of the radiation spectrum depend on the particular type of event. For instance, short GRBs have very hard photon spectra, while supernovae tend to have softer photon spectra. However, the impact of SNe is also determined by the longer-lived and more spread out (in time) cosmic ray flux. For any event (except those of a host star on its planets), distance is the key factor in determining total energy received. The overall luminosity (total emitted energy) varies with event type. Short GRBs, for instance, are less luminous than long GRBs, but also have harder spectra.

Estimates of rates of "dangerous" events depends then on the basic rate of occurrence in some chosen volume (e.g., a single galaxy) as well as the distance at which that event may have a serious impact on a biosphere, which again is determined by the total energy received (in turn determined by event luminosity and distance) and spectral hardness. Existing estimates have mainly been made considering impacts on an Earth-like planet, with depletion of $O_3$ as the main "dangerous" effect. This is likely oversimplified. First, some recent work has indicated that $O_3$ depletion associated with a GRB (and events with similar total energy received and spectral hardness) may not be as disastrous as previously thought, at least on certain primary producers in the oceans (Thomas et al., 2015, 2016). If correct, this reduces the rate of dangerous events, since it requires either more energetic or closer events, both of which would be less common. On the other hand, very recent work has shown that SNe may be more damaging through the extended high-energy cosmic ray flux, not so much through $O_3$ depletion but through irradiation by secondary particles (muons), and possibly through increased atmospheric ionization at very low altitudes, which may impact global climate (Thomas et al., 2016).

Estimates for the frequency of severe effects, using $O_3$ depletion as a measure of "severe," arrive at one dangerous event every few hundred million years for Earth for SNe and both types of GRBs, with SNe and short GRBs being slightly more frequent than long GRBs (Melott and Thomas, 2011). One could extend that to any Earth-like planet with oxygen-containing atmospheres, but the rates vary through cosmic time, as discussed below.

Of particular interest is the recent discovery that at least one, and probably several, core-collapse SNe occurred relatively near Earth a few million years ago (Fry et al., 2015; Thomas et al., 2016; Wallner et al., 2016). This has been very well established by geochemical evidence, but the distance to the SNe is large enough so that terrestrial effects were not very severe.

In all cases, it should be noted that "sterilization" of a habitat is an extreme condition. For every realistic event, refugia would exist in the deep ocean and under at least 100s of meters of ice or rock. While surface life may be dramatically affected and mass extinction may result, it is likely that some life would persist. In Earth's history, at least five major mass extinctions have occurred, including one that wiped out some 90% of species on Earth at the end of the Permian period. At least, one of these is statistically likely to have been connected with an astrophysical ionizing radiation event (a specific proposal has been made regarding the late Ordovician mass extinction, see Melott and Thomas, 2009). But in every case, life has returned and flourished. Therefore, talk of "sterilization" of planets is likely overblown except in the most extreme and rare of events.

On the other hand, when considering conditions in the universe before Earth's formation, such sterilization may be more realistic. In galaxies with very high star formation rate, planets formed within dense stellar areas could be exposed to intense and repeated supernova and even GRB events. A long-term exposure to very closeby events could indeed knock back or delay the development of life.

In addition, planets in the liquid-water habitable zone around low-mass stars may experience so much bombardment from stellar activity as to be stripped of their atmosphere which is quite likely to spell the end of any complex life there.

When considering the threats over cosmic time (the last 13 billion years), rate estimates need to take into account various factors. In particular, estimates of the rates of GRBs and SNe depend on star formation rate histories. Long GRBs and core-collapse SNe result from high-mass stars that are relatively short-lived (a few million years or so) and so track regions and periods of active star formation. Short GRBs require pairs of evolved objects such as neutron stars. These objects are generally considered to be the remnant of high-mass stars, and so depend in a similar way on star formation. Type Ia supernovae require a white dwarf, which is the remnant of a star with mass similar to that of the Sun or a few times higher. Such objects, then, require longer time periods to form, since a Solar lifetime is roughly 10 billion years. These events, then, will not directly track with active star formation. Simulations that track star formation and metallicity have been used to investigate where, as well as when, different regions of our own galaxy may have been habitable, as controlled by SNe and GRBs (see Gowanlock et al., 2011; Morrison and Gowanlock, 2015; Gowanlock, 2016). In general, they find that the inner part of the galaxy is more dangerous.

The picture for GRBs is complicated by the observation that long GRBs tend to occur in lower metallicity environments. This means that the long GRB rate would have been higher earlier in the universe's history. On the other hand, short GRBs do not show such a metallicity dependence. Recently, two groups have examined the role of long GRBs in the history of life in the Universe. Piran and Jimenez (2014) find that, due mainly to the metallicity dependence, the inner part of our galaxy is most dangerous and that the existence of life in any galaxy would be severely constrained by GRBs before about 5 billion years ago. If this is correct, then habitability before the rise of life on Earth may have been significantly limited by this kind of stellar explosion.

However, Li and Zhang (2015) come to a more optimistic conclusion, that about 50% of galaxies would be hospitable (considering only effects of GRBs) at about 9 billion years ago and 10% at about 11 billion years ago, and that the most hospitable galaxies are those similar to the Milky Way. These results make the earlier universe look much more likely to have been habitable, at least from the perspective of GRB threats. Li and Zhang (2015) also note that their results should be similar for SNe, though may not track exactly, since SNe do not have the same metallicity dependence as long GRBs.

Since AGN are powered by supermassive black holes at the centers of galaxies, there will be a "sweet spot" in cosmic history where they will be most active. First, enough time must have passed for the galaxy and its central black hole to form. Second, AGN appear to be active for some time and then become less active. This is likely due to the black hole clearing out material in the central part of its galaxy. Once most of the accessible matter has been consumed, the activity is likely to cease or at least become less intense and less sustained. In general, AGN are not thought to be a major constraint on habitability, except within the central regions of galaxies, which are already dangerous due to higher rates of SNe (Dartnell, 2011; Gobat and Hong, 2016; Dayal et al., 2016).

## 6. CONCLUSIONS

Here we have presented an overview of sources of biologically relevant astrophysical radiation, and effects of that radiation on organisms and their habitats. This chapter was focused on radiation as a constraint for habitability, due to the potential harmful effects of radiation on life "as we know it." Some of these effects have been known for a long time from studies of photobiology and radiobiology. The impact of radiation on life can be varied and complicated, and in some cases, by no means fully understood. From the astrobiological point of view, it is necessary to consider these effects in the context of astrophysical scenarios, which significantly may differ from the conditions of the present Earth. Even though some limitations may arise in reproducing or simulating these environments from the experimental point of view, these kinds of studies may provide an approximation of a real case scenario to estimate the probability of a planetary body to be habitable. Additionally, of particular interest is the potential for radiation to have positive effects, either for individuals or for the development and evolution of life. Some of them were briefly described in this chapter. This is an active area of research and it may well be that a future review such as this will find that radiation is as helpful as harmful, from the broad perspective of life in the universe.

Of necessity, our review does not cover all the details of particular impacts or the responses avail- able to organisms for dealing with radiation. We encourage the interested reader to follow-up with the sources cited for more details and to follow the continually changing landscape of this work. Surely, there is much more to be learned and we look forward to seeing what our community discovers over the next years and decades.

ACKNOWLEDGMENTS

BCT acknowledges support from NASA grant number NNX14AK22G under the Astrobiology: Exobiology and Evolutionary Biology program.

XCA acknowledges support from CONICET, Argentina.

2014. The effects of stellar winds on the magnetospheres and potential habitability of exoplanets. Astron. Astrophys. 570, A99. http://dx.doi.org/10.1051/0004-6361/201424323.

Segura, A., Krelove, K., Kasting, J.F., Sommerlatt, D., Meadows, V., Crisp, D., Cohen, M., Mlawer, E., 2003. Ozone concentrations and ultraviolet fluxes on Earth-like planets around other stars. Astrobiology 3, 689–708.

Segura, A., Walkowicz, L., Meadows, V., Kasting, J., Hawley, S., 2010. The effect of a strong stellar flare on the atmospheric chemistry of an Earth-like planet orbiting an M dwarf. Astrobiology 10, 751–771.

Setlow, P., 2006. Spores of Bacillus subtilis, their resistance to and killing by radiation, heat and chemicals. J. Appl. Microbiol. 101, 514–525.

Shahmohammadi, H.R., Asgarini, E., Terat, H., Saito, T., Ohyama, Y., Gekko, K., Yamamoto, O., Ide, H., 1998. Protective roles of bacterioruberin and intracellular KCl in the resistance of *Halobacterium salinarium* against DNA-damaging agents. J. Radiat. Res. 39, 251–262.

Sinha, R.P., Kumar, H.D., Kumar, A., Hlider, D.-P., 1995. Effects of UV-B irradiation on growth, survival, pigmentation and nitrogen metabolism enzymes in cyanobacteria. Acta Protozool. 34, 187–192.

Smith, D.S., Scalo, J., 2007. Solar X-ray flare hazards on the surface of Mars. Planet. Space Sci. 55, 517–527. http://dx.doi.org/10.1016/j

Smith, D.S., Scalo, J., Wheeler, J.C., 2004a. Transport of ionizing radiation in terrestrial-like exoplanet atmo- spheres. Icarus 171, 229–253. http://dx.doi.org/10.1016/j.icarus.2004.04.009.

Smith, D.S., Scalo, J., Wheeler, J.C., 2004b. Importance of biologically active aurora-like ultraviolet emission, stochastic irradiation of Earth and Mars by flares and explosions. OLEB 34, 513–532.

Tabataba-Vakili, F., Grenfell, J.L., Grießmeier, J.-M., Rauer, H., 2016. Atmospheric effects of stellar cosmic rays on Earth-like exoplanets orbiting M-dwarfs. Astron. Astrophys. 585, A96. http://dx.doi.org/10.1051/0004-6361/201425602.

Tarter, J.C., Backus, P.R., Mancinelli, R.L., Aurnou, J.M., Backman, D.E., Basri, G.S., Boss, A.P., Clarke, A., Deming, D., Doyle, L.R., Feigelson, E.D., Freund, F., et al., 2007. A reappraisal of the habitability of planets around M dwarf stars. Astrobiology 7, 30–65.

Thomas, B.C., Melott, A., Jackman, C., Laird, C., Medvedev, M., Stolarski, R., Gehrels, N.,

FURTHER READING

Cockell, C.S., 2000. Ultraviolet radiation and the photobiology of Earth's early oceans. OLEB 30, 467–499.

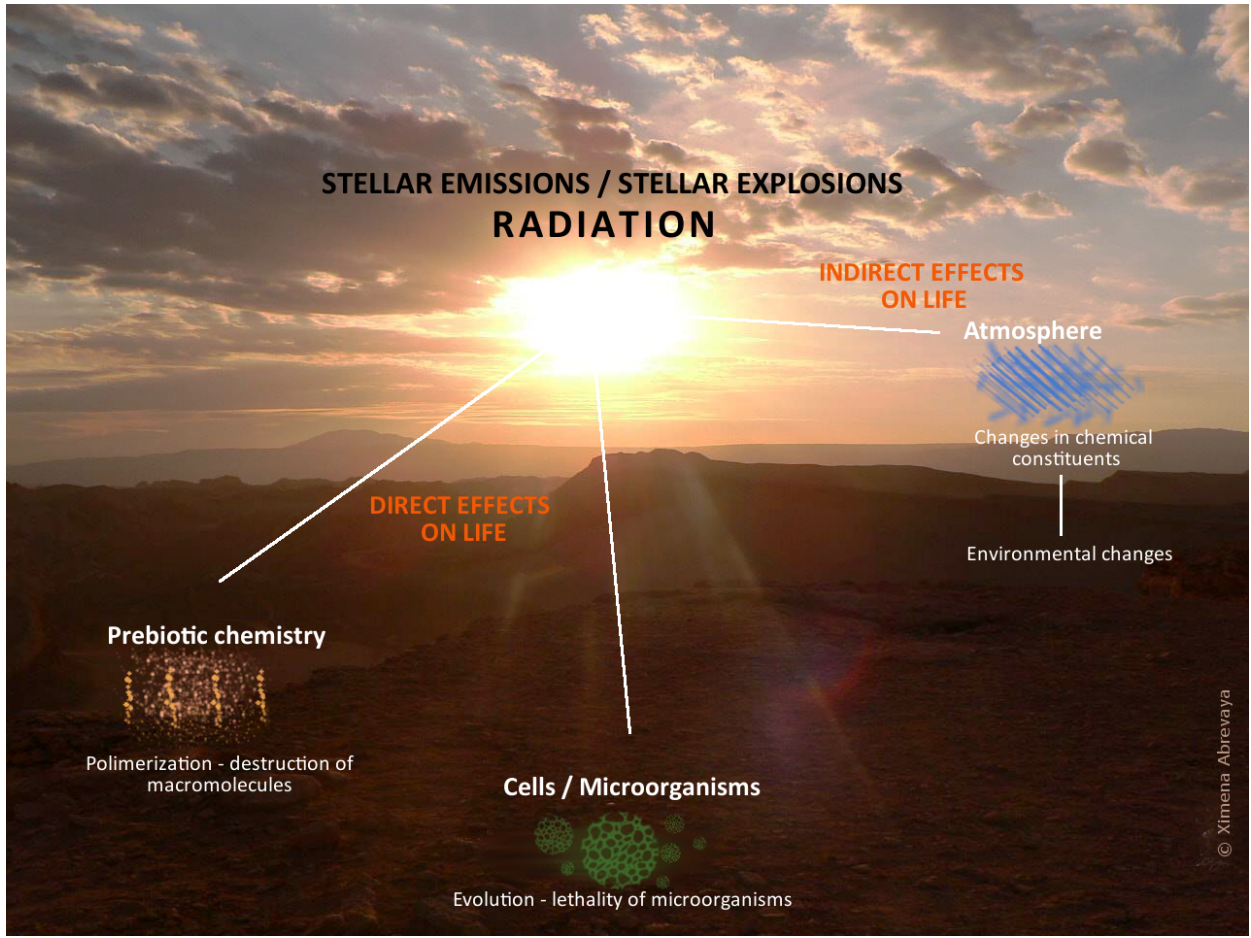

FIG. 1

Sources of radiation and its impact on life through direct and indirect effects.

(Modified from Abrevaya 2013)